# Self-organized criticality in living systems

C. Adami

W. K. Kellogg Radiation Laboratory, 106–38, California Institute of Technology

Pasadena, California 91125 USA

(December 20,1993)

## Abstract

We suggest that ensembles of self-replicating entities such as biological systems naturally evolve into a self-organized critical state in which fluctuations, as well as waiting times between phase transitions ("epochs"), are distributed according to a $1/f$ power law. We demonstrate these concepts by analyzing a population of coexisting self-replicating strings (segments of computer code) subject to mutation and survival of the fittest.

Typeset using REVTEX

Self-organized criticality [1] is the term generically applied to systems that drive themselves to a critical state that is robust to perturbations and whose macroscopic behavior is predictable to the extent that it follows power laws with exponents that depend only on geometry and the spatial structure. In general, the microscopic processes giving rise to the self-organized critical state are dissipative transport processes associated with a threshold, or critical, variable. The paradigm for the self-organized critical state is the sandpile: the critical state is the self-similar and robust pile itself, distribution of sizes and duration of avalanches (resulting from perturbations) follow distinct power laws, and grains of sand are transported if the local slope of the pile exceeds a critical value, thus restoring the critical state.

It has been suggested [2–6] that biological populations are typically in a self-organized critical state, evidenced for example by a power-law distribution of extinction events. Furthermore, it was observed that population structures gleaned from taxonomic data [7] show a fractal geometry. While this is a very appealing idea, especially in view of the robustness of living systems, it has suffered from being somewhat vague, mainly because of the difficulty involved in modeling living systems. Specifically, there is as yet neither a clear identification of the self-organized critical state of life or the agent that causes self-organization, nor a definition of a critical or threshold variable whose disturbance causes the ubiquitous avalanches giving rise to power-law distributions.

In this Letter, we report the observation of self-organized criticality in an artificial living system, the tierra environment [8,9]. In this system, strings of machine language-like instructions with the ability to self-replicate in core memory "live" and co-evolve in an environment subject to random mutation and selection of the fittest. As such, it is not a *simulation* of life but rather *artificial* life. Interestingly, the system displays some of the uncanny hallmarks known from simple proto-cellular systems. Most importantly, this artificial environment offers the chance to control the microscopic processes leading to complex behavior, such as replication and mutation. Furthermore, the macroscopic behavior is predictable via the usual methods of statistical mechanics applied to an ensemble of self-replicating entities.

Simple equations [9] reveal that in this system the fitness of a string is determined by its replication rate, as measured by executing the string's instructions (its "genome") and counting the number of offspring per unit time. The principle of "survival-of-the-fittest" boils down to a "survival-of-the-most-populous." Fitness is then a quantity that is *genotype specific*, i.e., each individual arrangement of instructions in a string translates into a specific replication rate. This is a unique feature that makes this system fundamentally different from the molecular quasi-species model of Eigen *et al.* [10]. We can thus think of fitness as a highly complicated function on the space of all strings, while the population is characterized by the current average value. This state, however, is metastable: a successful mutation can create a new "best" genotype (or "master sequence" [10]) with a higher replication rate that disrupts the equilibrium and induces a phase transition to a new "vacuum," defined by the new dominant master sequence and its offspring.

Let $\epsilon_i$ stand for the replication rate of genotype $i$, and $\langle \epsilon \rangle$ for its average over the population. In a mean-field approximation, for strings of length $\ell$ subject to a mutation rate $R$ (we are considering here only external "cosmic-ray" mutations, which have an effect similar to copy-errors), the critical variable is the growth factor

$$\gamma_i = \epsilon_i - \langle \epsilon \rangle - R\ell . \tag{1}$$

In the equilibrium situation, the master sequence and its $\epsilon$-degenerate offspring and mutants have $\gamma_i = 0$, while inferior species have $\gamma_i < 0$. This prevents exponential growth of the most successful species in the long run. An advantageous mutation, however, can make $\gamma_i > 0$ for the new master sequence. Such a disturbance causes the information contained in the master sequence to be transmitted throughout the system via the offspring, giving rise to avalanches that are scale-independent. Gradually, all genotypes with a subcritical replication rate will become extinct and be replaced: the system returns to its critical state. Clearly, the normal state of such a population of self-replicating entities is a superposition of a very large number of metastable states, with transitions between them induced by mutation and copy-errors. It undergoes spontaneous phase transitions if a mutation creates a genotype with $\gamma_i > 0$, ushering in a new "epoch" of domination by a new species. In fact, in such a system there is no scale that would set the average time between avalanches, nor is there a scale setting the size of the avalanche. The latter is determined by the amount of information gained by the new master sequence. We thus expect both distributions to be given by power laws.

To test this hypothesis, we have analyzed the evolution of an ensemble of strings subject to a Poisson-random mutation rate of $R = 0.5 \times 10^{-8}$ mutations per site per unit time (the unit of time is the execution of one instruction). The strings are segments of computer code of a specially developed instruction set with only 32 instructions running on a virtual computer. A mutated instruction most likely will cause the program to "break," yet occasionally may improve it. The strings live in a strip of memory with a total number of 131,072 sites that can represent one instruction each, with periodic boundary condition (i.e., the strip wraps on itself). It is typically inhabited by 600–1400 strings of length 60–150, all offspring of a single (handwritten) progenitor that is able to self-replicate and used to "inoculate" the strip. For purposes of reproducibility, we used the original "ancestor" written by Tom Ray [8], who created the tierra system just described. As described elsewhere [9], this ancestor is well-suited for evolutionary experiments due to the amount of redundancy in its code. The fitness landscape that this population explores is determined by all possible ways to reduce the time to gestate a single offspring (the gestation time) and the opportunity to trigger bonus CPU time by developing the "genetic code" necessary to perform certain user-specified tasks (see Ref. [9] for details on this environment). In other words, we provide an environment containing information that the strings can discover (through adaptive mutation) and exploit.

The population adapts to this environment through discontinuous jumps, as evidenced in Fig. 1.

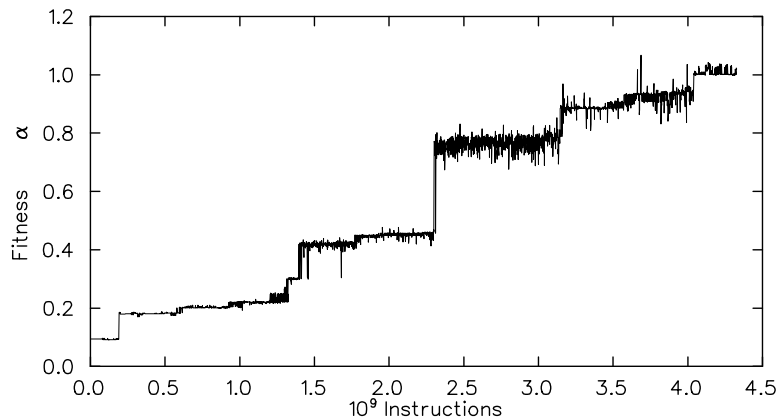

FIG. 1 Fitness curve for a typical run. The fitness parameter $\alpha$ of the most successful (i.e., most populous) genotype is plotted as a function of time, measured every million instructions for a mutation rate $R = 0.5 \times 10^{-8}$.

There, we have plotted the "fitness-of-the-best" $\alpha$ versus total number of instructions executed (i.e., time elapsed) every million instructions for a typical run. For technical reasons, the measured quantity $\alpha$ is the replication rate $\epsilon$ multiplied by the total number of instructions allocated to the strings in one "sweep" through the population, with $0 \leq \alpha \leq 1$. The latter bound is imposed only to maintain parallelism: in order to emulate parallel coexistence, each string is allocated a certain slice of CPU time and executed serially (see, e.g., Ref. [8] for details). The visible noise in Fig. 1 is mainly due to mutations and finite-size effects. This noise, of course, drives the fitness jumps that adapt the population to the environment.

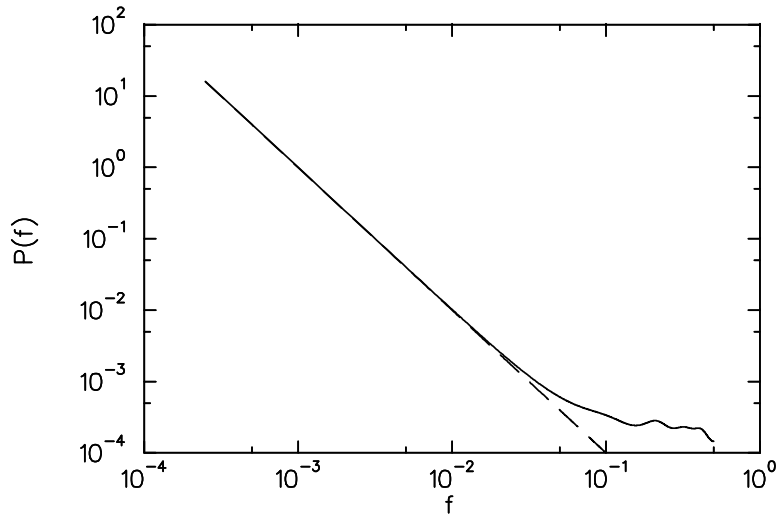

FIG. 2 Power spectrum $P(f)$ of a typical fitness curve $\alpha(t)$ (Fig. 1). The dashed line is a fit to $P(f) \sim f^{-\beta}$ with $\beta = 2.0 \pm 0.05$.

We have plotted in Fig. 2 the power spectral density of a typical fitness history, which reveals a clear power-law distribution with $P(f) \sim f^{-\beta}$ and $\beta = 2.0 \pm 0.05$. Scaling exponents from other runs are compatible within the error bars quoted.

Fluctuations distributed according to a power law are the telltale sign of a self-organized critical state [1]. To further our understanding of this state we have also measured the distribution of waiting times between phase transitions, or length of epochs, in 50 runs under identical conditions (save the random number seed), resulting in 512 measured waiting times.

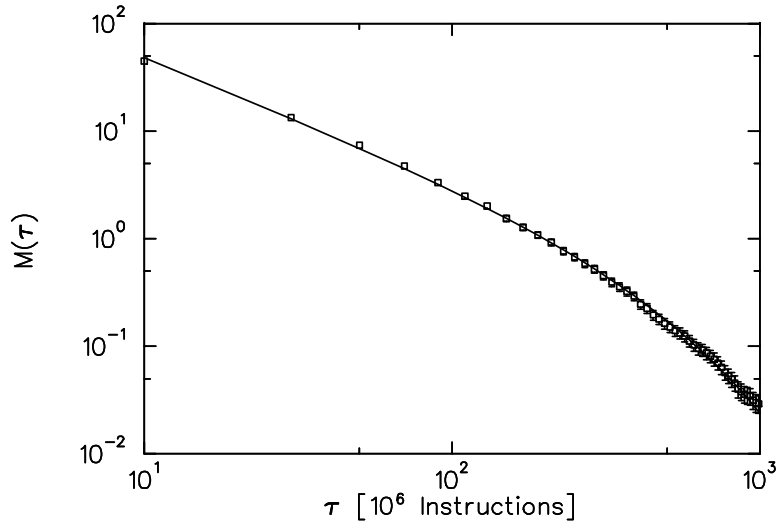

FIG. 3  Integrated distribution of times between phase transitions $\tau$ (length of epoch). The solid line is a fit to the data with $M(\tau) \sim \tau^{-\alpha} \exp(-\tau/T)$ with $\alpha = 1.1 \pm 0.05$ and a cut-off parameter $T = 450 \pm 50$ modeling finite-size effects.

Figure 3 shows the waiting time $\tau$ plotted versus the *integrated* distribution function

$$M(\tau) = \frac{1}{\tau} \int_\tau^\infty N(t) dt , \qquad (2)$$

where $N(t)$ is the distribution function. The integrated quantity is distributed with the same exponent as $N(t)$, but is more reliable even with poor statistics on $N(t)$. To obtain the waiting times, we determined that a phase transition occurred if the fitness jumps discontinually to a new level with a fitness increase of a minimum of 7.5%. As the resulting plot proves that the fitness curves are fractal, this condition cannot change the power law. Rather, fitness curves like the one in Fig. 1 are expected to look similar *at all scales*. Clearly, as we cannot measure waiting times $t \geq 500$ [11] with good statistical accuracy due to the finite lengths of our runs, the integrated distribution function shows finite-size effects that we model with a cut-off parameter $T$. Thus, we fit

$$M(\tau) = c\,\tau^{-\alpha} \exp\left(-\frac{\tau}{T}\right) \qquad (3)$$

and find $\alpha = 1.1 \pm 0.05$ and $T = 450 \pm 50$. The critical exponents reported here do not allow for an unambiguous determination of the universality class of the model reported here, of the effective dimension of the space occupied by the strings. A manifestly two-dimensional version of the model and a mean-field theory describing it are in preparation.

We have suggested that the normal state of an ensemble of self-replicating entities is self-organized criticality, the agent of self-organization being information. We identified the growth factor $\gamma_i$ as the critical variable and described avalanches of "invention" that drive the adaptation of the population. We tested these hypotheses in the artificial life system **tierra** and found power-law distributions in the power spectrum of fitness fluctuations, as well as in the distribution of waiting times. Self-organized criticality in living systems has wide-ranging consequences for theories of evolution. On the one hand, gradualism is incompatible with criticality, and a punctuated equilibrium picture is favoured (see, e.g., Ref. [12]). On the other hand, the fractal nature of the fitness history (Fig. 1) would account for fitness improvements on all scales driven only by microscopic mutations.

Note added: After completion of this manuscript, we noticed the appearance of Ref. [13], wherein conclusions similar to ours are drawn from a simple evolutionary model.

## ACKNOWLEDGMENTS

This work was supported in part by the National Science Foundation, Grant No. PHY90-13248.

# REFERENCES

[1] P. Bak, C. Tang, and K. Wiesenfeld, Phys. Rev. Lett. **59**, 381 (1987); Phys. Rev. A **38**, 364 (1988).

[2] M. Eigen, Adv. Chem. Phys. **33**, 211 (1978).

[3] D. M. Raup, Science **231**, 1528 (1986).

[4] S. A. Kauffman and S. Johnsen, J. Theor. Biol. **149**, 467 (1991).

[5] S. A. Kauffman, *The Origins of Order* (Oxford University Press, 1993).

[6] P. Bak, Physica **A 191**, 41 (1992).

[7] B. Burlando, J. Theor. Biol. **146**, 99 (1990); **163**, 161 (1993).

[8] T. S. Ray, in *Artificial Life II: A Proceedings Volume in the Santa Fe Institute in the Sciences of Complexity, Vol. 10*, edited by C. G. Langton (Addison-Wesley, Reading, Massachusetts, 1992).

[9] C. Adami, *Learning and Complexity in Genetic Auto-Adaptive Systems*, Caltech preprint MAP-164 (1993), to appear in *Complex Systems*.

[10] M. Eigen, J. McCaskill, and P. Schuster, Adv. Chem. Phys. **75**, 149 (1989).

[11] Waiting times are measured in units of millions of instructions executed.

[12] S. J. Gould and N. Eldredge, Nature **366**, 223 (1993).

[13] P. Bak and K. Sneppen, Phys. Rev. Lett. **71**, 4083 (1993).